\documentclass[apj]{emulateapj}
\usepackage{apjfonts}

\def\chandra    {\emph{Chandra}}
\def\xmm        {\emph{XMM}}
\def\vla        {\emph{VLA}}
\def\rosat        {\emph{ROSAT}}
\def\gmrt {\emph{GMRT}}
\def\aspcs {Astronomical Society of the Pacific Conference Series}
\def\phr {Physics Reports}
\def \deg      {$^{\circ}$}

\def\arcsec{$^{\prime\prime}$}

\def\lax{\lesssim}

\shorttitle{Shock front in A754}
\shortauthors{G.~Macario et al.}

\begin{document}

\setlength{\pdfpageheight}{\paperheight}
\setlength{\pdfpagewidth}{\paperwidth}

\slugcomment{submitted to {\em The Astrophysical Journal}}

\title{A shock front in the merging galaxy cluster Abell 754:
\\
X-ray and radio observations }

\author{Giulia Macario\altaffilmark{1,2}, 
Maxim Markevitch\altaffilmark{2},
Simona Giacintucci\altaffilmark{1,2},
Gianfranco Brunetti \altaffilmark{1}, \\
Tiziana Venturi\altaffilmark{1},
Stephen S. Murray\altaffilmark{2}
}
\altaffiltext{1}{INAF/Istituto di Radioastronomia, via Gobetti 101, 
I-40129, Bologna, Italy; g.macario@ira.inaf.it}
\altaffiltext{2}{Harvard-Smithsonian Center for Astrophysics, 
60 Garden Street, Cambridge, MA 02138, USA}

\begin{abstract}

We present new \chandra\ X-ray and \textit{Giant Meterwave Radio Telescope} (\gmrt) 
radio observations of the nearby merging galaxy cluster
Abell 754. Our X-ray data confirm the presence of a shock front by obtaining  
the first direct measurement of a gas temperature jump across the 
X-ray brightness edge previously seen in the imaging data. 
A\,754 is only the fourth galaxy cluster with confirmed merger shock fronts, 
and it has the weakest shock of those, with a Mach number 
M$=1.57^{+0.16}_{-0.12}$. In our new \gmrt\ 
observation at 330 MHz, we find that the 
previously-known centrally located radio halo extends eastward to the
position of the shock. The X-ray shock front also coincides with the
position of a radio relic previously observed at 74 MHz.  The radio spectrum
of the post-shock region, using our radio data and the earlier results at 
74 MHz and 1.4 GHz, is very steep.  We argue that acceleration 
of electrons at the shock front directly from thermal to ultrarelativistic
energies is problematic due to energy arguments, 
while reacceleration of preexisting relativistic electrons is more plausible.

\end{abstract}



\section{Introduction}
\label{sec:intro}

Galaxy clusters form via mergers of smaller substructures.  During such
mergers, most of the kinetic energy of the gas belonging to the colliding
subclusters is dissipated by shocks and turbulence into thermal energy of
the intracluster medium (ICM) of the resulting system.  Shocks and
turbulence are also expected to amplify the cluster magnetic fields and
accelerate cosmic ray particles from the thermal ICM, or reaccelerate
preexisting relativistic particles.  These non-thermal components manifest
themselves as diffuse synchrotron radio sources, known as radio halos and
relics (e.g., Ferrari et al.\ 2008; Cassano\ 2009 for recent reviews), and 
inverse Compton X-ray emission at high energies
(e.g., Fusco-Femiano et al.\ 2004; Rephaeli \& Gruber 2002; see, however,
Wik et al.\ 2009).

Shock fronts represent a unique observational tool to study the physical
processes in the ICM. They create sharp discontinuities in the cluster X-ray
surface brightness images and allow to measure the gas velocities in the sky
plane using X-ray imaging spectroscopy (e.g., Markevitch, Sarazin, \&
Vikhlinin 1999). Shock-heated regions are routinely observed in merging
clusters. However, known shock {\em fronts}\/ are still rare, because one
has to catch the front when it has not yet moved to the outer, low surface
brightness regions of the cluster, where the background X-ray emission
dominates.  Moreover, the merger has to be occurring nearly in the plane of
the sky, otherwise projection effects could hide the gas density and
temperture jumps.

Until now, reliable detections of merger shock fronts have been reported in
only three galaxy clusters. One is the Bullet cluster (1E\,0657--56,
Markevitch et al.\ 2002), another is Abell 520 (Markevitch et al.\ 2005),
and the two fronts have been recently discovered in Abell 2146 (Russell et
al.\ 2010).  Only in these clusters both the sharp gas density edges and the
unambiguous temperature jumps were found, allowing the identification of the
brightness feature as a shock and determination of the gas velocity.  In
this paper, we report on the new shock front in the merging cluster Abell 754. 
We present the analysis of \chandra\ observations and study the connection 
between the shock and the diffuse cluster radio emission, through 
{\it Giant Metrewave Radio Telescope} (\gmrt) observations 
at 330 MHz and {\it Very Large Array} (\vla) archival data at 1.4 GHz.

We assume a flat cosmology with H$_0$ = 70 km s$^{-1}$ Mpc$^{-1}$ and
$\Omega_{0}$ = 0.3, in which 1$^{\prime\prime}$ is 1.054 kpc at the redshift
of A754. We adopt the convention $S \propto \nu^{-\alpha}$ for the synchrotron 
spectrum. 
Uncertainties are 68\%, unless stated otherwise.


\section{The merging cluster A754}
\label{sec:a754}

A754 is a rich nearby cluster at $z=0.0542$ (Struble \& Rood 1999) in the
stage of a violent merger.  It has been actively studied in the optical and
X-ray bands, and is considered the prototype of a major cluster merger.
Previous studies revealed that the cluster has a complex galaxy distribution
(Fabricant et al.\ 1986, Zabludoff \& Zaritzki 1995), X-ray morphology, and
gas temperature structure (Henry \& Briel 1995; Henrikesen \& Markevitch
1996; Markevitch et al.\ 2003, hereafter M03; Henry et al.\ 2004).  These
data indicate that A754 is undergoing a major merger of two components 
along an east-west axis, probably with a non-zero impact parameter. 
Hydrodynamic simulations were able to reproduce most of the observed 
X-ray features in A754 by considering an off-axis collision between two 
subclusters with a mass ratio of 2.5:1 (Roettiger et al.\ 1998).  
\chandra\ data suggested that the merger may be more complex, 
possibly involving a third subcluster or a cloud of cool gas decoupled 
from its former host subcluster (M03).

Krivonos et al.\ (2003) reported an edge-like surface brightness feature in
the \rosat\ PSPC image of the cluster, located east of the core, which
looked like a shock front. They modeled it with a radial density profile
with a jump, deriving a Mach number $M=1.71^{+0.45}_{-0.24}$ from this jump
under the assumption that this is indeed a shock front.  This brightness
feature was also apparent in the \chandra\ image by M03, but it was at the
edge of the \chandra\ field of view, which did not allow its detailed study.
Henry et al.\ (2004) presented a temperature map of the cluster from an
\xmm\/-\textit{Newton} observation.  The presumed post-shock region 
was shown to be hot, as expected for a shock front.  However, 
because the X-ray brightness ahead of this front is very low, 
the \xmm\ observation did not have sufficient sensitivity to determine 
the temperature in the pre-shock region, in order to detect the expected 
temperature jump.  They were only able to derive a 
temperature in the annulus around the cluster that included the pre-shock
region (region 12 in Fig.\ 6 of Henry et al. 2004), but that annulus is dominated
by emission from regions on the other side of the cluster, unrelated to the
shock front. 
We note that many features that look like shocks in the X-ray
images turned out to be ``cold fronts'' 
(Markevitch \& Vikhlinin 2007), and the difference between these phenomena
is the sign of the temperature jump at the feature.  In this work, we
present a direct temperature measurement in the pre-shock region from a new
\chandra\ observation.

A754 also exhibits evidence for ultrarelativistic electrons and magnetic
fields that coexist with the ICM.  Fusco-Femiano et al.\ (2003) reported a
3$\sigma$ hard X-ray excess at $E>45$ keV with \emph{BeppoSAX}, which may
come from inverse Compton scattering of the Cosmic Microwave Background (CMB) 
photons on relativistic electrons.  
A clear evidence of nonthermal emission in A754 comes from the
radio observations.  
A radio halo was detected at 74 MHz, 330 MHz and 1.4 GHz with the \vla\, 
(Kassim et al.\ 2001, hereafter K01; Bacchi et al.\ 2003, hereafter B03, 
and references therein). Moreover, the presence of two peripheral 
radio relics were reported in K01, East and West of the radio halo. 
Only the eastern emission was seen in B03. 
The whole cluster diffuse radio emission has been recently studied in the 
frequency range 150--1360 MHz by Kale et al.\ (2009).


\section{Chandra observations and the shock front}
\label{sec:chandraobs}

A754 was first observed with \chandra\ ACIS-I in October 1999 (OBSID 577).
This pointing was centered on the cluster center and used by M03 to study
the cluster temperature structure.  In Febraury 2009, a new long \chandra\ 
ACIS-I observation (OBSID 10743) was pointed on the putative shock front, in
order to derive the temperature profile across the shock.  The new exposure
partially overlaps the old one (see Fig.\ \ref{fig:xray}), and in this paper
we use both observations for the spectral analysis of regions where they
overlap.

Both datasets were processed and cleaned in a standard manner (e.g.,
Vikhlinin et al.\ 2005).  The data were cleaned of flares and the blank-sky
datasets were normalized as described in Markevitch et al.\ (2003b).  OBSID
577 had small flare contamination and the final clean exposure was 39 ks. No
background flares were present in the new observation, so we used the full
exposure of 95 ks.  The old observation was taken in FAINT mode, while the
new one in VFAINT, which allowed additional background filtering to be
applied.  The detector + sky background was modeled using the blank-sky
background datasets corresponding to the dates of observations (periods B
and E, respectively).  We normalized the blank-sky backgrounds using the the
ratio of counts in the high-energy band 9.5--12 keV, which is free from the
sky emission. This correction was within 10\% of the exposure ratios, as
expected.

All point sources were masked out for the extraction of brightness profiles
and spectra.  The instrument responses for spectral analysis were generated
weighting the detector ARF and RMF with the cluster brightness within each
spectral extraction region (Vikhlinin et al.\ 2005). We used the most recent
calibration products --- namely, version N0008 for the telescope effective
area, N0006 for the CCD quantum efficiency and gain files, and N0005 for the
ACIS time-dependent low-energy contaminant model. We also assessed the
systematic uncertainty of our results by using a newer, experimental update
to the ACIS contaminant model (A. Vikhlinin, private communication; see
below).

Figure \ref{fig:xray} shows a slightly smoothed ACIS image of the cluster in
the 0.5--4 keV band, obtained by co-adding the two observations and
correcting for exposure nonuniformity.  As found by previous studies, the
cluster undergoes a complex merger with the main axis along the NW-SE
direction.  The image clearly shows a brightness edge to the east of the
dense elongated core, perpendicular to the merger direction.  This is 
the putative shock front reported by Krivonos et al.\ (2003). 
The new observation provides
sufficient statistics to determine the exact nature of this feature.

\subsection{Density profile across the edge}
\label{sec:d_profile}

\begin{figure*}
\epsscale{1.12}
\plotone{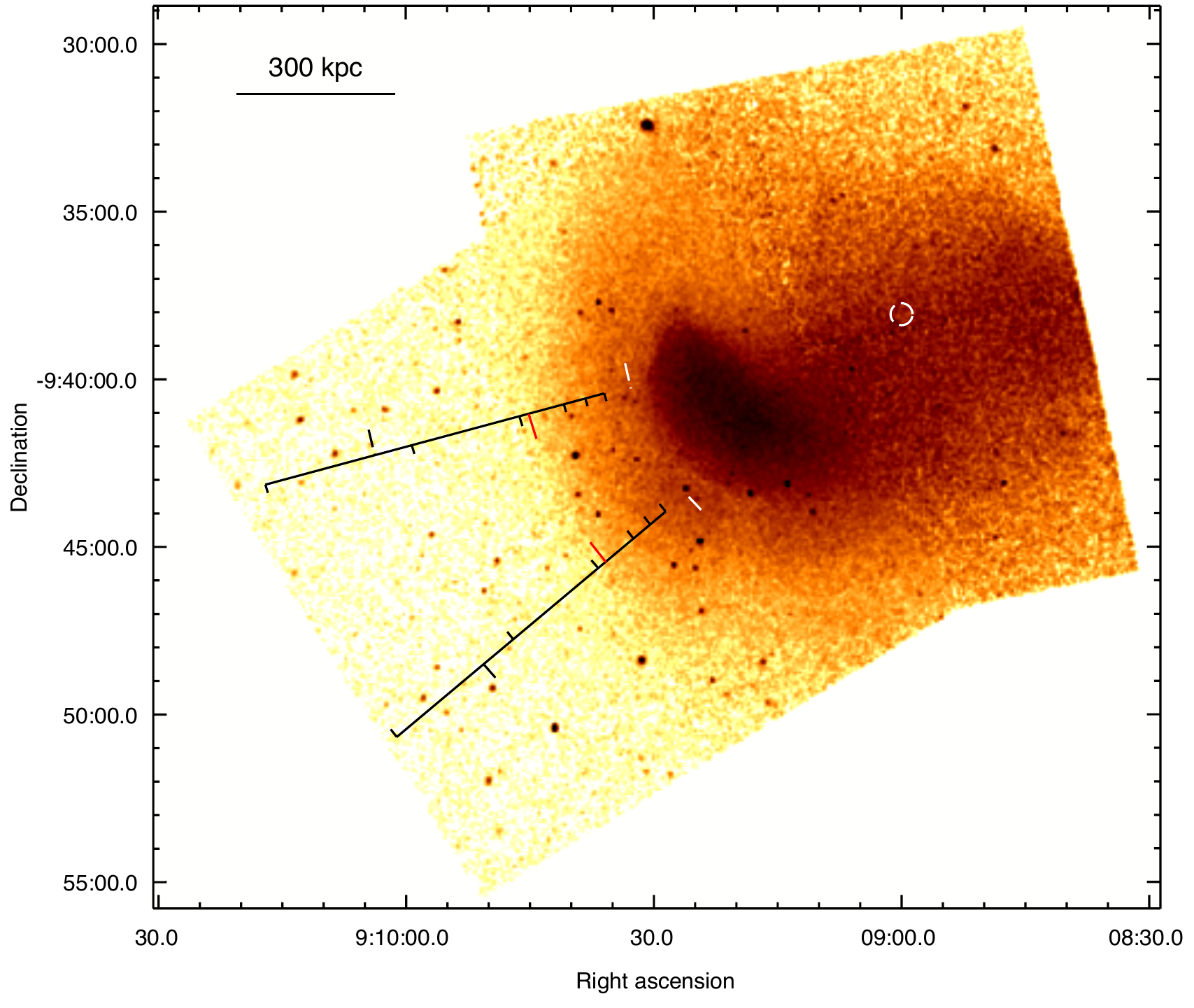}
\caption{\chandra\, slightly smoothed 0.5--4 keV image of A754 from a sum
  of the new 95 ks exposure and the archival 39 ks observation. Also shown is the 25\deg\ 
  sector used to extract the X-ray surface brightness profile and for the spectral analysis (see \S\ \ref{sec:d_profile} and \S\ \ref{sec:t_profile} ). Dashed ticks outside the sector mark the radial range used for the 
  profile fitting; small ticks inside the sector mark the radial bins chosen to derive the temperature profile; bigger red ticks indicate the best--fit density jump, measured from its center of curvature (marked by the dashed circle).}
\label{fig:xray}
\end{figure*}

\begin{figure*}[t]
\epsscale{1.} 
\plotone{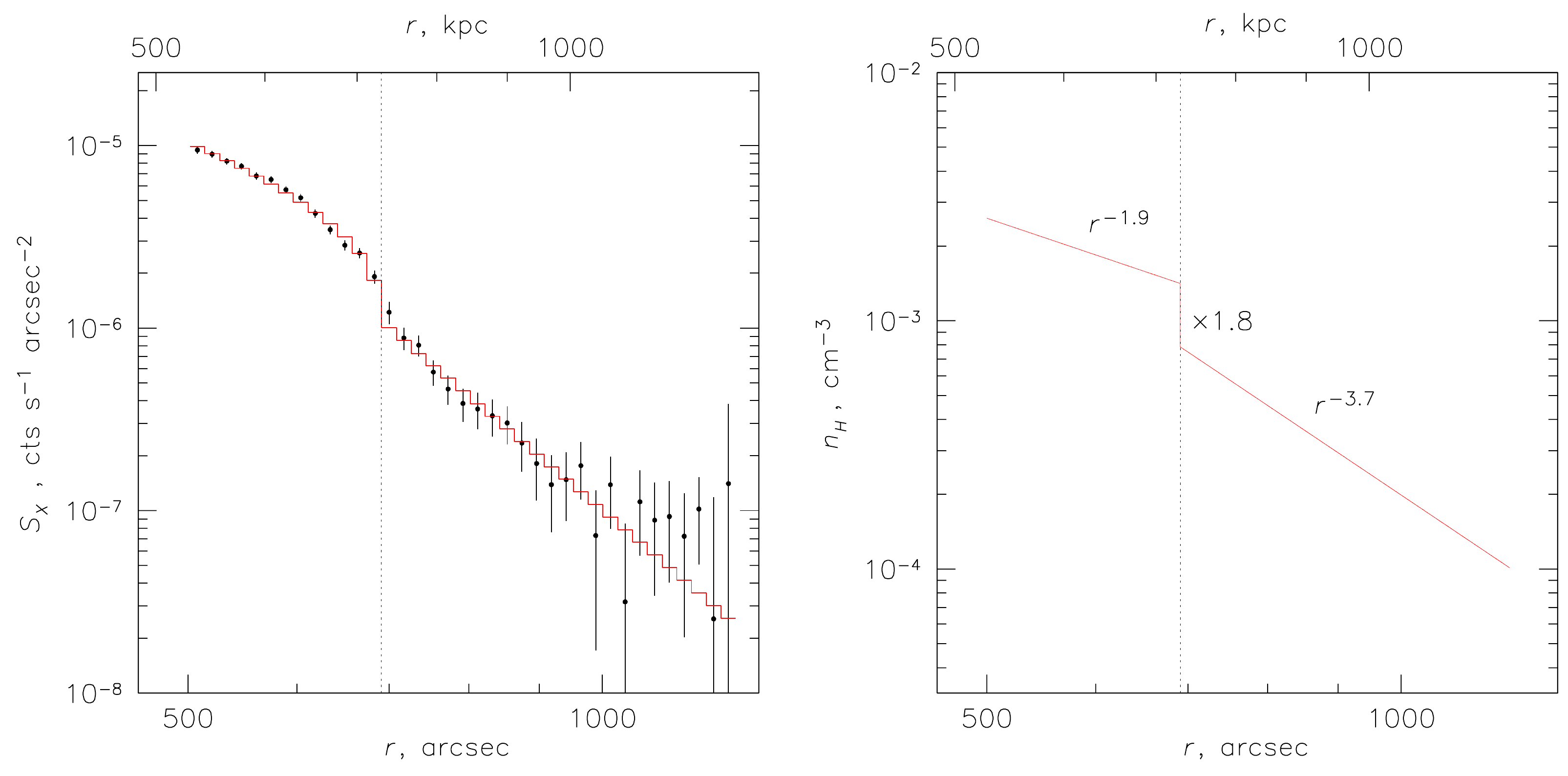}
\caption{({\em a}) The 0.5-4 keV surface brightness profile across the front
  in the sector shown in Fig.\ \ref{fig:xray} (see text). Errors are
  $1\sigma$; histogram shows the best-fit model that corresponds to the
  radial density profile shown in panel ({\em b}). The dashed line marks the
  best-fit position of the shock front (measured from its center of
  curvature).}
\label{fig:profiles}
\end{figure*}

Figure \ref{fig:profiles} shows a radial X-ray surface brightness profile
across this edge, extracted in a 25\deg\ sector shown in Fig.\ 
\ref{fig:xray}. To facilitate proper geometric modeling, the sector is
centered on the center of curvature of the brightness edge 
(dashed white circle in Fig.\ \ref{fig:xray}) and encompasses
its most prominent segment (avoiding the apparent decrease of the edge
brightness contrast at greater opening angles).  All point sources are
carefully masked.  The energy band for the profile was restricted to 0.5--4
keV to minimize the dependence of X-ray emissivity on temperature and
maximize the signal-to-noise ratio. Only OBSID 10743 is used
for the brightness profile, since the earlier observation does not cover the
pre-shock region.  The exposure of this observation alone is more than
sufficient for the post-shock region. We consider
only the region outside the bright elongated core, which is
obviously unrelated to the shock.

The brightness profile across the edge (Fig.\ \ref{fig:profiles}a) exhibits
the typical shape of a projected spherical density
discontinuity (Markevitch et al.\ 2000), and we fit it with
such a model, under the assumption that the curvature along
the line of sight is the same as in the image plane. The
model radial density profile that we use consists of two
power laws, $\rho\propto r^{\alpha_1}$ and $\rho\propto
r^{\alpha_2}$, on two sides of the edge, with an abrupt jump
at the edge (Fig.\ \ref{fig:profiles}b).  For each set of the profile
parameters, we projected the corresponding emission measure
profile onto the image plane and fit to the observed X-ray
brightness profile (Fig.\ \ref{fig:profiles}a), treating the slopes and the
radius and amplitude of the jump as free parameters. The fit
was restricted to the interesting radial range --- staying
as close as possible to the edge on the inside in order to
avoid contamination from unrelated core structure, but
extending as far as possible on the outside for proper
deprojection (see dashed ticks on the outer sides 
of the sector in  Fig.\ \ref{fig:xray}).  As seen
from Fig.\ \ref{fig:profiles}a, the model fits very well. The best-fit
density jump, after a 3\% correction of the $0.5-4$ keV
Chandra brightness for the measured temperature difference
across the edge (see \S\ \ref{sec:t_profile} below), is $\rho_2/\rho_1 =
1.80^{+0.20}_{-0.15}$. The confidence interval is evaluated
by allowing all other model parameters to be free.


%
\begin{table}
\caption[]{X-ray temperature fits}
\begin{center}
\renewcommand{\arraystretch}{1.4}
\renewcommand{\tabcolsep}{2.2mm}
\footnotesize
\begin{tabular}{p{1.6cm}ccc}
\hline
\hline
Region & obsid 577 & obsid 10743 & Simult.\ fit \\
\hline
\multicolumn{4}{c}{Current (N0005) ACIS contaminant model:}\\
1 \dotfill& $17.6^{+6.6}_{-4.4}$&$~9.0^{+1.0}_{-0.7}$& $10.1^{+1.0}_{-0.9}$\\
2 \dotfill& $~9.6^{+2.2}_{-1.7}$&$14.2^{+2.5}_{-2.1}$ &$12.5^{+1.9}_{-1.8}$\\
3 \dotfill& $~8.7^{+2.2}_{-1.5}$&$16.4^{+3.7}_{-2.8}$ &$13.3^{+2.5}_{-1.9}$\\ 
4 \dotfill& ...                 &$~7.4^{+2.3}_{-2.0}$ & ... \\
5 \dotfill& ...                 &$~4.5^{+2.4}_{-1.5}$ & ... \\
3 deproj \dotfill& ...         & ...                  &$16.4^{+5.1}_{-3.5}$\\
\hline
\multicolumn{4}{c}{Experimental ACIS contaminant model:}\\
3 \dotfill& $~8.7^{+2.2}_{-1.5}$&$14.2^{+2.8}_{-2.4}$ &$12.2^{+2.1}_{-1.8}$  \\
4 \dotfill& ...                 &$~6.6^{+2.3}_{-1.5}$ & ... \\
\hline
\end{tabular}
\end{center}
\label{tab:tfits}
\end{table}

\begin{figure}[t]
\epsscale{1.15}
\plotone{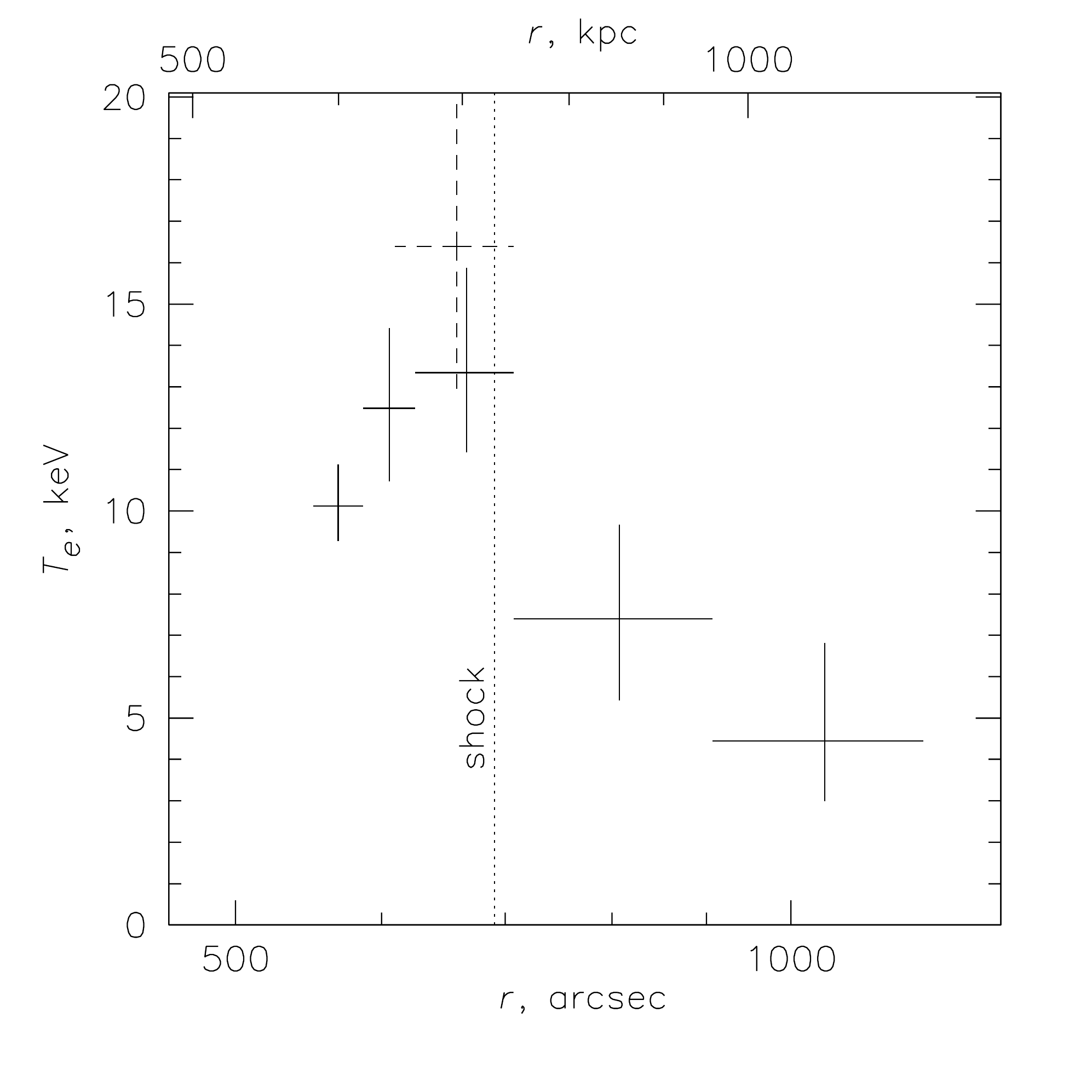}
\caption{Temperature profile across the shock (errors are 68\%). Solid crosses
  show projected temperature fits; dashed line for the first post-shock
  region shows a deprojected temperature (modeling the projected
  contribution of the pre-shock gas based on the best-fit density model).
  For post-shock regions, simultaneous fits for two observations are shown,
  while two outer regions are covered only by the recent observation (see
  Table \ref{tab:tfits} for more details).  Dotted vertical line shows the
  position of the shock front from the density fit (see Fig.\ 
  \ref{fig:profiles}).}
\label{fig:tprofile}
\end{figure}

\subsection{Temperature profile across the edge}
\label{sec:t_profile}

We now extract spectra from several radial bins in the same sector across
the density edge (Fig.\/\ref{fig:xray}; \S\/\ref{sec:d_profile}) to determine the gas temperature
jump across the edge. There is enough statistical accuracy to divide the
low-brightness region outside the edge into two bins, and the inner brighter
region into three bins, avoiding getting to close to the cluster cool
elongated core. The five radial bins are marked by ticks inside the sector 
(Fig.\/\ref{fig:xray}); the longer red ticks mark the best--fit position of the 
density jump (Fig.\/\ref{fig:xray}; \S\/\ref{sec:d_profile}). 
The first high-brightness region adjacent to the jump
extends across the best-fit jump position a bit, in order to avoid
contaminating the adjacent low-brightness region by any irregularities in
the edge shape.  The spectra were fit in the 0.8--9 keV band with XSPEC,
using the absorbed thermal plasma model WABS(APEC), with elemental
abundances as free parameters, fixing the absorption column to the Galactic
value ($N_H = 4.82 \times 10^{20}$ cm$^{-2}$, Kalberla et al.\ 2005). We
also tried to free $N_H$, and all fits were consistent with the Galactic
value. All fits had acceptable $\chi^2$ values.  The systematic uncertainty
of background modeling was evaluated by varying the blank-sky background
normalization by $\pm 2$\% (68\%, Hickox \& Markevitch 2005) and added in
quadrature to the statistical uncertainties. It was negligible for the
brighter regions inside the edge, but similar to the statistical uncertainty
for the two low-brightness bins (the observation exposure was selected to
achieve this).

Solid crosses in Fig.\ \ref{fig:tprofile} show the resulting temperature
values, using the simultaneous fit to both observations for the inner
regions and OBSID 10743 for the outer regions. The figure shows a clear
temperature jump of the sign that corresponds to a shock front.

Table \ref{tab:tfits} gives separate fits for two pointings; the regions are
numbered in order of increasing radius in Fig.\ \ref{fig:tprofile}. We note
that two \chandra\ observations are in mild disagreement for the post-shock
regions (though none deviates by more than $2\sigma$ from the simultaneous
fit).  This may be caused by residual calibration problems --- these regions
are at the edge of the ACIS-I field of view in OBSID 577, while at the
center in OBSID 10743. In addition, these observations are separated by 10
years.  We tried an experimental update to the recently released time
dependence of the time-dependent ACIS-I contaminant model (A.  Vikhlinin,
private communication) to see if it changes the results qualitatively, for the two
bins adjacent to the front.  The results are shown in Table \ref{tab:tfits}.
The disagreement is slightly reduced. Most importantly, the temperature jump
does not change qualitatively, and is also seen in the recent observation
alone, using either calibration. To the extent the results from different
spatial regions can be compared, our post-shock high temperatures are in
broad agreement with those derived with \xmm\ by Henry et al.\ (2004).

Thus, the temperature profile confirms that the
brightness edge is a shock front. From the Rankine-Hugoniot
jump conditions, we can use our best-fit gas density jump to
derive a Mach number of this shock. Assuming monoatomic gas
with $\gamma = 5/3$, we obtain $M=1.57^{+0.16}_{-0.12}$

The temperature jump can give an independent Mach number estimate, though
usually with a lower accuracy. To check for the consistency of these
estimates, we first get a deprojected temperature in the first post-shock
radial bin, since in the spherical geometry, the temperature we measure
there is affected by projection of the cooler emission from the outer,
pre-shock bins. (At the same time, the pre-shock temperature profile appears
consistent with isothermal, so the true temperature should be close to the
projected one, to a sufficient accuracy).  From the best-fit density model,
we estimate a projected emission measure fraction in bin 3 from the outer
regions 4 and 5 in Fig.\ \ref{fig:tprofile} to be 21\% and 3\%, respectively.
Neglecting the latter and fitting the spectrum for region 3 using an
additional thermal model with the normalization and temperature (including
its uncertainties) corresponding to region 4, we obtain a ``deprojected''
temperature shown by the dashed cross in Fig.\ \ref{fig:tprofile}.

The ratio of the post-shock to pre-shock temperatures is thus
$2.2^{+1.1}_{-0.7}$, which corresponds to $M=2.1^{+0.7}_{-0.6}$ (68\%), 
consistent (within its expected larger uncertainties) with the value
from the density jump obtained above.


\section{Radio observations}
\label{sec:radioobs}

In this section, we present our \gmrt\ observations of A754 at 330 MHz, 
and analysis of the archival \vla\ data at 1.4 GHz.  Details of observations
are summarized in Table \ref{tab:obs}, which reports the telescope, project code, 
observing date, frequency, total bandwidth, total time on source,
synthesized half-power bandwidth (HPBW) of the full array, rms level
(1$\sigma$) at full resolution, $u-v$ range, and the largest detectable
structure (LDS).

%
\begin{table*}[htbp!]
\caption[]{Summary of the radio observations.}
\begin{center}
\footnotesize
\begin{tabular}{cccclccccc}
\hline\noalign{\smallskip}
\hline\noalign{\smallskip}
Radio  &  Project & Observation & $\nu$ & $\Delta \nu$  & t  & HPBW, p.a. & rms & {\it u-v} range & LDS  \\
  telescope & code & date & (MHz)& (MHz) & (min) & ($^{\prime
     \prime}\times^{\prime \prime}$, $^{\circ}$)&  ($\mu$Jy beam$^{-1}$)
   & (k$\lambda$) & ($^{\prime}$)   \\
\noalign{\smallskip}
\hline\noalign{\smallskip}
 \gmrt & 08GBA01 &June 23, 2005 & 325 & 32(16)$^{*}$ &  150 & 10.0$\times$9.1,
 -64 & 450
  & $\sim$ 0.08--25 &  $\sim 32$ \\
 \vla-D & AF372 &Sept. 25, 2000 & 1365/1435$^{**}$ & 50 &  160 &
 64.5$\times$38.8, 5 & 50 & $\sim$ 0.13--4.7 & $\sim 15$\\
\noalign{\smallskip}
\hline\noalign{\smallskip}
\end{tabular}
\end{center}
Notes to Table \ref{tab:obs}: $^{*}$ Only the USB dataset was used for the 
analysis (see Sect. \ref{sec:gmrtobs}). $^{**}$ Only the the lower
frequency IF was used for the analysis (see Sect. \ref{sec:vlaobs}).
\label{tab:obs}
\end{table*}
%

\subsection{\gmrt\/ observations at 325 MHz}
\label{sec:gmrtobs}

A754 was observed with \gmrt\ at 325 MHz in June 2005 for a total time on
source of 2.5 hr (Table \ref{tab:obs}). The observations were carried out
using the upper and lower side bands simultaneously (USB and LSB,
respectively), for a total observing bandwidth of 32 MHz. The default
spectral-line observing mode was used, with 128 channels for each band and a
spectral resolution of 125 kHz/channel. The LSB dataset was corrupted and
could not be processed, so only the USB data,  
which are centered at approximately 330 MHz, 
were used to produce the images presented here.
The dataset was calibrated and analyzed using the NRAO
Astronomical Image Processing System package (AIPS). We refer to Giacintucci
et al.\ (2008) for a complete description of the data reduction procedure 
used in this paper. 

Due to the large field of view of \gmrt\ at 330 MHz (primary beam $\sim
1.8^{\circ}$), we used the wide-field imaging technique at each step of the
phase self-calibration process, to account for the non-planar nature of the
sky. We covered a field of $\sim 2.7^\circ\times 2.7^\circ$ with 25 facets.
The final images were produced using the multi-scale CLEAN implemented 
in the AIPS task IMAGR, which results in better imaging of extended sources 
compared to the traditional CLEAN (e.g., Clarke \& Ensslin 2006; for a detailed 
discussion, see Appendix A in Greisen, Spekkens, \& van Moorsel 2009). 
We used three circular Gaussians as model components. 
One of the Gaussian was chosen to have zero width to 
accurately model point sources and small-scale structures, 
and the other two have a width of 20$''$ and 45$''$ respectively, 
to progressively highlight the extended emission during the clean.

Beyond the image at full resolution ($10.0'' \times 9.1''$), we produced
images with lower resolution (down to $\sim 100''$), tapering the $u-v$ data
by means of the parameters {\tt robust} and {\tt uvtaper} in the task IMAGR.
Even though only half of the data was usable, the sensitivity of the final
images is quite good --- the rms noise level (1$\sigma$) ranges from 0.45 
mJy beam$^{-1}$ in the full resolution image to $\sim$1 mJy beam$^{-1}$
in the lowest resolution images. We estimate that the flux density
calibration uncertainties are within $5\%$.

\subsection{\vla\/ archive data at 1.4 GHz}
\label{sec:vlaobs}

\vla\ observations of A754 at 1.4 GHz (project AF372) were presented by B03.
We extracted these observations from the archive and re-analyzed them in
order to ensure the best possible comparison with our \gmrt\ data.  
The observations were obtained using the D
configuration and two IFs, centered at 1365 MHz and 1435 MHz (see
Table~\ref{tab:obs} for details on the observations). 
Standard calibration and imaging were carried out using AIPS.  The dataset
was self-calibrated in phase only.  The higher frequency IF was found to be
affected by strong radio interference, which compromises the quality of the
images produced using both IFs. For this reason, only the low-frequency IF
was used for the analysis presented in this paper. The final images were
obtained implementing the multi-scale clean option in IMAGR, 
as for the 330 MHz data (see \S\ref{sec:gmrtobs}).
\\
The rms noise in the image at full resolution is 50
$\mu$Jy beam$^{-1}$ (Table~\ref{tab:obs}).  A slightly higher noise
(1$\sigma \sim 60-70$ $\mu$Jy beam$^{-1}$ was achieved in the low-resolution
(HPBW$\sim$70$^{\prime \prime}$) images.   
The average residual amplitude errors in the data are of the order
of $\lax 5$\%.


\begin{figure*}
\begin{center}
\epsscale{1.15}
\plotone{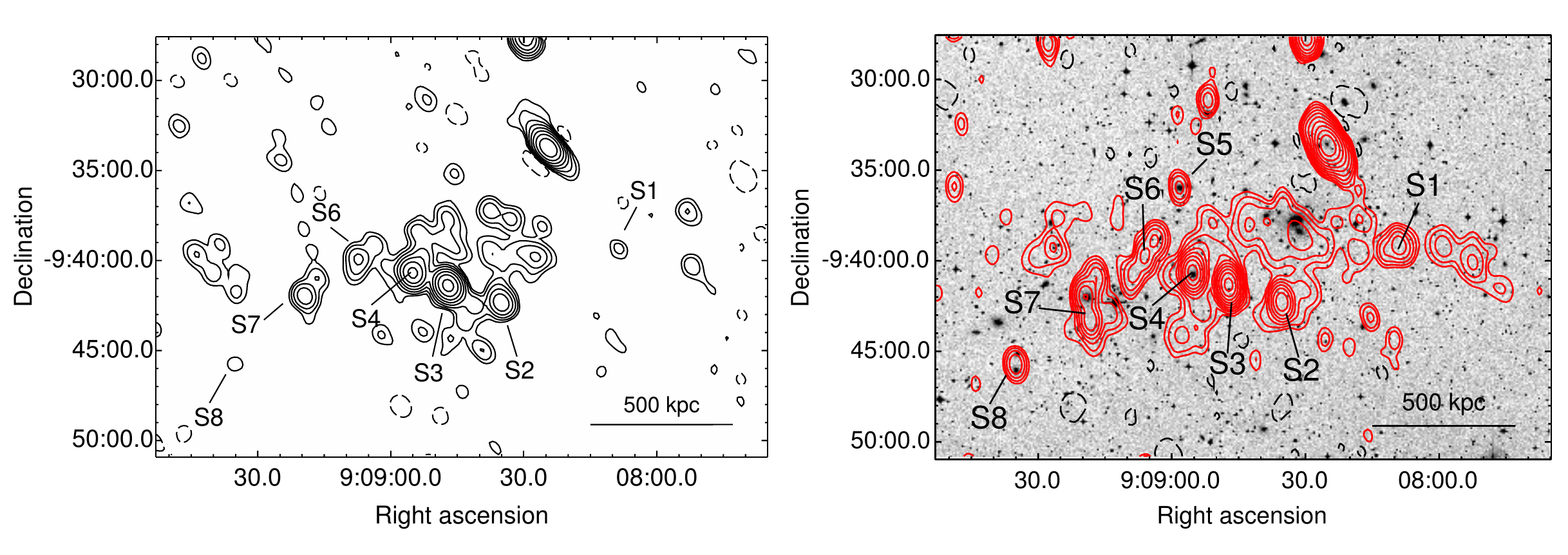}
\caption{{\it Left}: \gmrt\, 330 MHz image of A\,754 at the resolution
of 63.5$^{\prime\prime} \times 52.1^{\prime\prime}$,
p.a. $30^{\circ}$. The  1$\sigma$ level is 1.1 mJy beam$^{-1}$. Contours start
at $-3.3$ mJy beam$^{-1}$ (dashed) and $+3.3$ mJy beam$^{-1}$, and
then scale by a factor of 2. 
{\it Right}: \vla\, full-resolution image at 1365 MHz (contours),
overlaid on the optical POSS-2 red image of A\,754. 
The restoring beam is $64.5^{\prime\prime} \times 38.8^{\prime\prime}$,
p.a. $5^{\circ}$, and the 1$\sigma$ noise level is 50 $\mu$Jy beam$^{-1}$.  
Contours are spaced by a factor of 2, starting from $\pm$0.15 mJy
beam$^{-1}$ (negative contours are shown as dashed). In both images, labels
indicate the discrete radio galaxies identified at 1.4 GHz by B03.}
\label{fig:gmrt_vla}
\end{center}
\end{figure*}

\begin{figure*}
\epsscale{1.15}
\plotone{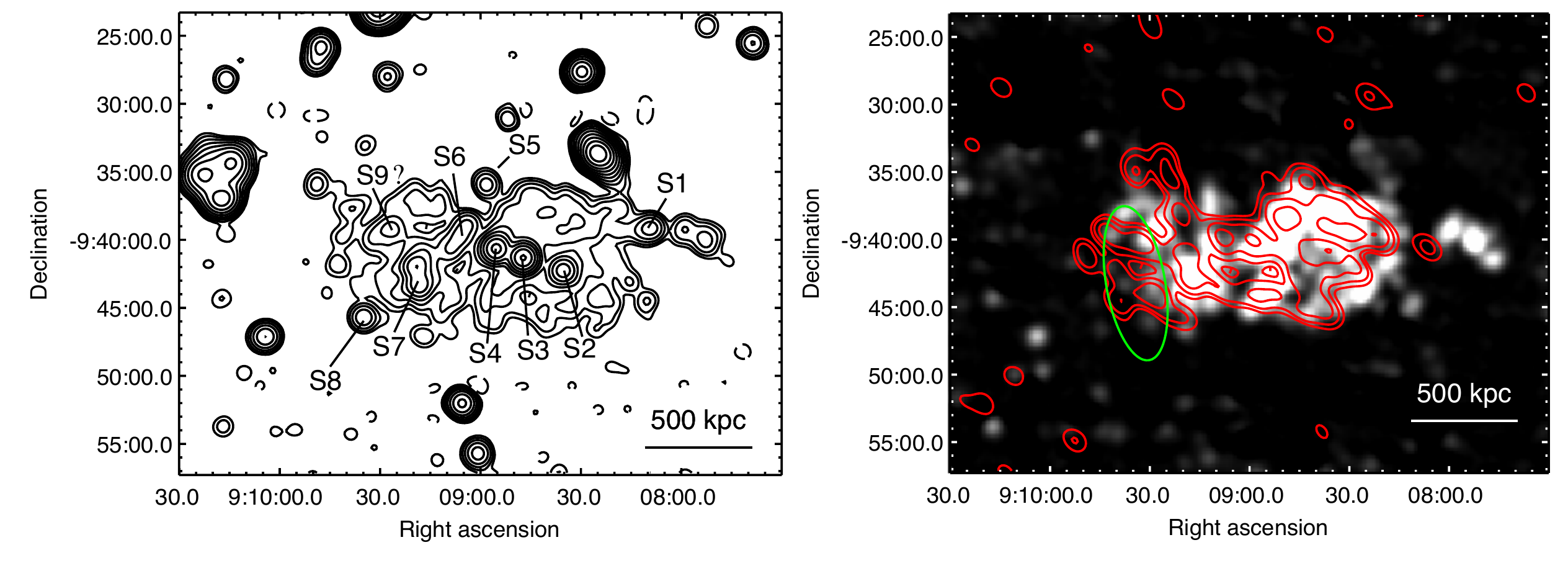}
\caption{{\it Left}: \vla\, low-resolution image at 1365 MHz. 
The restoring beam is $70''\times 70''$, p.a. 0; contours are 
spaced by a factor 2 starting from $\pm 3 \sigma$, the 1$\sigma$ level is 60  $\mu$Jy beam$^{-1}$. 
Labels indicate the discrete radio galaxies identified within the diffuse emission. 
Negative contours are shown as dashed. 
Ticks mark contours that correspond to holes in the radio brightness. 
{\it Right}: 
diffuse emission after the subtraction of
all identifiable discrete sources. Red contours show a
low-resolution \gmrt\/ image at 330 MHz with a
restoring beam of $109''\times 74''$, p.a. $43^\circ$ and
the $1\sigma$ noise level of 1 mJy beam$^{-1}$. Contours are
spaced by a factor of 2, starting from $3\sigma$. Greyscale
shows a low-resolution \vla\/ image at 1365 MHz ---
same as in left panel but without the discrete sources.  The
peak flux density is 3.77 mJy beam$^{-1}$ and the lowest
level of greyscale corresponds to $\sim 2\sigma$. The green
ellipse is the area covered by the $2\sigma$ contour (0.4 Jy
beam$^{-1}$) of the radio relic in the \vla\/ image at
74 MHz (K01). 
Ticks mark contours that correspond to holes in the radio brightness. 
}
\label{fig:a754_multifreq}
\end{figure*}

\begin{figure*}[ht!]
\epsscale{0.95}
\plotone{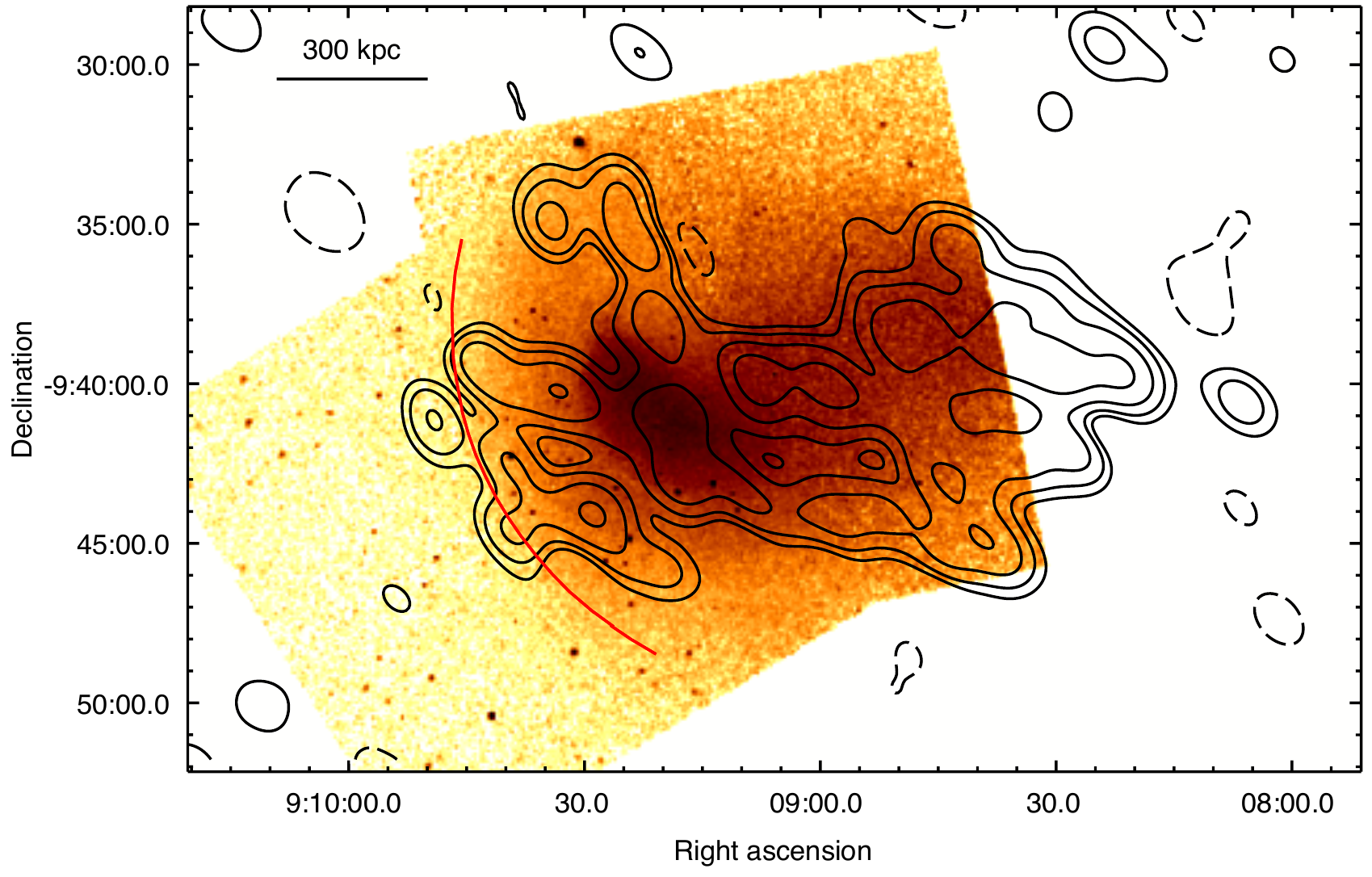}
\caption{\gmrt\, low resolution radio contours (same as in Fig.\ \ref{fig:a754_multifreq}, right panel),  overlaid on the \chandra\, image of A754 (same as Fig. \ref{fig:xray}, see \S\ \ref{sec:chandraobs}). The dashed contours show the 3$\sigma$ negative values; the red arc indicates the best fit position of the density jump (see \S\ \ref{sec:d_profile}).
}
\label{fig:radiox2}
\end{figure*}


\section{Radio analysis}\label{sec:radio_an}

In Fig.\ \ref{fig:gmrt_vla}, we present the new
\footnote{These observations, available from the \gmrt\,  public archive,  
has been also recently used by Kale et al. 2009, for their low frequency 
study of A\,754. However, they do not show the radio images obtained from these data.} 
\gmrt\ image at 330 MHz
at the resolution of $63''\times 52''$ (left panel) and the \vla\ 1365 MHz
image at full resolution (HPBW=$64''\times 39''$), overlaid on the optical 
POSS-2 red frame of A\,754 (right panel). 
The first positive contour corresponds to the 3$\sigma$ level in both images 
(i.e., 3.3 mJy beam$^{-1}$ at 330 MHz and 0.21 mJy beam$^{-1}$ at 1365 MHz).   
Numerous discrete radio sources are located in the cluster area. 
Some of these sources are embedded within the diffuse emission. 
We identify all the radio galaxies found  by B03. 
Following their same notation, these are labelled from S1 to S8 in both images. 
The point source S5 is undetected at 330 MHz.

In order to properly image the cluster diffuse emission, we subtracted  
all the discrete radio galaxies in the cluster region from the $u-v$ data at both frequencies. 
We first subtracted the brightest radio sources in the field 
(including the extended radio galaxy north--east of S1 in Fig.\ \ref{fig:gmrt_vla}), to 
simplify the detection and subtraction of the fainter discrete galaxies  
embedded in the diffuse emission (from S1 to S8). 
\\
We finally produced a set of low-resolution images from the "subtracted"
data sets at both frequencies, to highlight the extended emission only.

Our results are shown in Fig.\ \ref{fig:a754_multifreq}. 
The left panel  shows contours of our \vla\, 
low-resolution image at 1365 MHz (HPBW=$70'' \times 70''$), before the subtraction of discrete 
radio sources
\footnote{
Here S9 ? indicate the peak of emission visible to the north--east of S7 in Fig.\ \ref{fig:gmrt_vla}. 
This apparently coincides with an optical counterpart, and might be an additional point source; 
however we considered it as a peak of the diffuse emission and we 
did not subtract it from the data.
}. 
The rms noise of the image is 60 $\mu$Jy beam$^{-1}$ (1 $\sigma$);
this is significantly lower with respect to the noise level reached by B03 in their image of same resolution (see Fig. 3 of B03). 
In the right panel of Fig.\ \ref{fig:a754_multifreq} we present images of the 
diffuse emission only, obtained after the subtraction of discrete sources described above. 
The \gmrt\ 330 MHz intensity contours at
the resolution of $109'' \times 74''$ are overlaid on the \vla\ 1365 MHz
image (grayscale), produced with a restoring beam of $70'' \times 70''$.
Both images show the very large extent and complexity of the diffuse
emission in A754. The diffuse source spans $\sim$ 1.4 Mpc in the east-west
direction and $\sim 750$ kpc along the north-south axis, as measured from
the 3$\sigma$ contour level at 330 MHz. 
The 1365 MHz image reveal a ridge in the radio brightness,  
 extended westbound the 330 MHz contours. 
 On the other hand, the high frequency emission is less extended in the eastern 
 region (see below). 
We point out that these are reliable features, not due to a shift between the two images. 
Indeed we carefully checked the astrometry of our radio images at both frequencies, 
by comparing the positions of the peaks of  the radio sources with that of the optical counterparts. 

The green ellipse in Fig.\ \ref{fig:a754_multifreq} (right panel) represents the
2$\sigma$ contour (0.4 Jy beam$^{-1}$) of the eastern radio relic found in the 74
MHz \vla\ image by K01.  
The 330 MHz diffuse emission extends eastward
to the outer side of this relic. 
Interestingly, we notice that the shock front found in \S\ \ref{sec:t_profile} coincides 
with the edge of the emission at 330 MHz (see Fig.\ \ref{fig:radiox2}), 
possibly suggesting a connection between the radio edge and the shock (see \S\ \ref{sec:discussion}) .
\\
We detect significant emission at 1365 MHz only in part of the region covered 
by the green ellipse (see Fig.\ \ref{fig:a754_multifreq}).
Our \vla\, images at 1365 MHz show more diffuse emission, south of the S2-S4 sources, 
with respect to B03.
This is most likely due to differences in the editing of the data, in the imaging procedure 
(multi-scale clean, see \S\ \ref{sec:vlaobs}) and in the accuracy of the subtraction of 
discrete sources (see also the comment in B03);  our final rms value is indeed a factor 1.6 
better.
\\
We do not detect any radio emission in the region of the western relic found
by K01.

\subsection{Spectrum of the diffuse radio emission}
\label{sec:radio}

The radio images in Fig.~\ref{fig:a754_multifreq} (right panel) show a complex, Mpc-scale
diffuse source associated with A754.  We use our 330 MHz \gmrt\ and the
reprocessed 1365 MHz \vla\ data presented in \S\ref{sec:radioobs}, along
with the \vla\ image at 74 MHz kindly provided by N. Kassim (see Fig.\ 1 in
K01), to derive its integrated spectrum. 
The flux densities of the diffuse emission, after the subtraction of the
discrete sources, are 828$\pm$41 mJy and 89$\pm$5 mJy at 330 MHz and 1365
MHz, respectively. 
These are also reported in Table\ \ref{tab:fluxes}, 
along with the associated uncertainties and angular
resolution of the images used for the measurements. 
We obtained these values by integrating over the same area  
that encompasses the whole diffuse emission as detected at both frequencies
\footnote{We also checked that no significant change in the spectral index 
occurs by integrating within the emission at the 2$\sigma$ level.}. 
We note that the flux density at 1365 MHz is lower than the value reported
 in B03 where, however, the authors notice that the
 subtraction of discrete sources is not accurate (B03; see also \S\ \ref{sec:radio_an}). 
The flux density of the whole diffuse emission at 74 MHz is
$\sim 6.6$ Jy (Tab.\ \ref{tab:fluxes}). This value was integrated over the same area described
above and does not include the contribution of the brightest radio
galaxies at the cluster center (i.e., S2, S3 and S4, and the extended radio galaxy 
north--east of S1; see Fig.\ \ref{fig:gmrt_vla}) 
which amounts to a total of $\sim$ 2.9 Jy (as estimated in K01). However, due to
the lower sensitivity, only part of the extended emission visible at 330
MHz and 1365 MHz is detected at 74 MHz. Thus, our estimate should be
considered as a lower limit to the flux. 
The spectral index for the whole diffuse emission between 
330 and 1365 MHz is $\alpha = 1.57 \pm 0.05$. 
This value is steeper than typically found for giant radio halos 
($\alpha \sim 1.2-1.3$; e.g., Ferrari et al.\ 2008), although a number of halos with
ultra-steep spectrum ($\alpha > 1.5$) has been discovered in the past few
years (e.g., Brunetti et al.\ 2008, Brentjens 2008, Macario et al.\ 2010). 

For simplicity, in the following we will call \textit{relic} the feature discovered by K01 at 74 MHz, 
that they defined as eastern radio relic (see also discussion, \S\ \ref{sec:discussion}).
\\
In our \gmrt\ image at 330 MHz, the
relic does not appear as a distinct feature separated from the radio halo,
nor a clear increment of the surface brightness of the diffuse emission is
detected in that region.  
This may imply that the spatial segregation between the relic and the halo in the 74 MHz image
(see Fig. 1 in K01) may be due to a steeper spectrum of the relic at $\nu \lax 300$  MHz 
combined with the lower sensitivity of the 74 MHz observation. 
\\
To obtain the radio spectrum of the relic region, 
we integrated the flux densities at 330 MHz
and 1365 MHz over the area defined by the 2$\sigma$ contour level of the
relic at 74 MHz (green ellipse in Fig.~\ref{fig:a754_multifreq},  right panel). We
obtained 106$\pm$5 mJy at 330 MHz and 6.0$\pm$0.3 mJy at 1365 MHz (see Table\ \ref{tab:fluxes}). 

\begin{figure}[ht!]
\epsscale{1.05}
\plotone{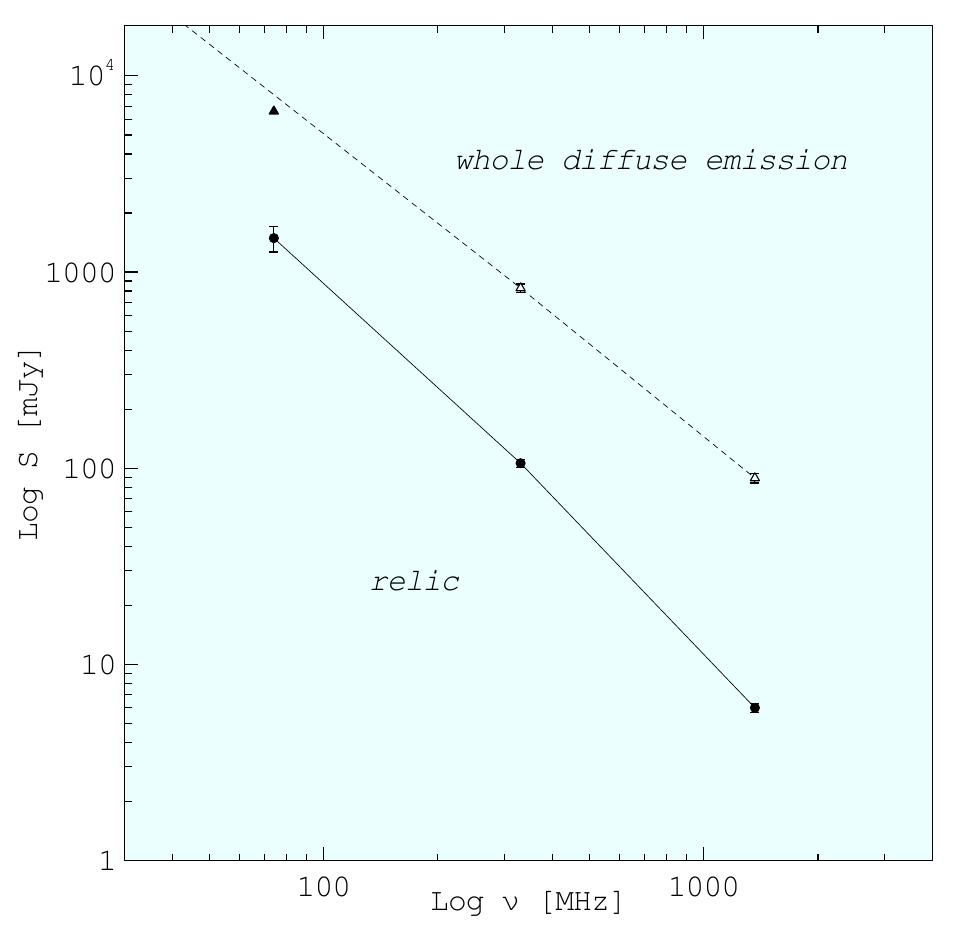}
\caption{
Integrated spectrum of relic region (filled dots) in A\,754 in the frequency range 74--1365 MHz. 
The solid line connects the values of flux density reported in Table \ref{tab:fluxes}. 
The spectrum of the whole diffuse emission between 330 and 1365 MHz is also shown (open triangles 
and  dashed line); the estimated flux density at 74 MHz (filled triangle) is in reasonable agreement with the extrapolation of the spectrum.
}
\label{fig:a754_edge_spec}
\end{figure}

In Fig.\ \ref{fig:a754_edge_spec} we show the integrated radio spectrum of the relic, 
in the frequency range 74--1365 MHz.   
The relic has indeed a very steep spectrum, with 
$\alpha^{0.3 GHz}_{0.07 GHz} = 1.77\pm0.10$, 
and $\alpha^{1.4 GHz}_{0.3 GHz} = 2.02\pm0.04$. 
An extrapolation of this power law to lower frequencies predicts a 74 MHz flux 
density that is consistent with that reported in K01. 
For a direct comparison, 
the flux densities of the whole diffuse emission are also shown. 

A spectral study of the diffuse emission in A\,754 was presented in Kale et al.\ (2009). 
However their analysis is limited to specific regions of emission detected 
in their \gmrt\, 150 MHz image (that do not show the whole extended emission). 
Therefore, we cannot compare our spectral results to those found by them.  

\begin{table}[htbp]
\caption[]{Flux densities of the diffuse radio emission in A\,754}
\begin{center}
\begin{tabular}{ccccc}
\hline\noalign{\smallskip}
\hline\noalign{\smallskip}
$\nu$ (MHz)    & HPBW , \arcsec $\times$ \arcsec    &  \multicolumn{2}{c}{ S$_{\nu}$ (mJy)} & Ref. \\
\cline{3-4}
   			&  								     &  \textit{relic}  & \textit{whole emission} &  \\   
\hline\noalign{\smallskip}
74     	&  $316 \times 232$		& 1489$\pm$223 &  $\sim$6600		&  this work; K01   \\
330        & $109 \times 74$ 		&  106$\pm$5 	& 828$\pm$41	& this work; Fig.\ \ref{fig:a754_multifreq} \\
1365    	& $70.0 \times 70.0$ 		&  6.0$\pm$0.3 	& 89$\pm$5		& this work; Fig.\ \ref{fig:a754_multifreq} \\
\hline\noalign{\smallskip}
\end{tabular}
\end{center}
\label{tab:fluxes}
\end{table}


\section{Discussion}
\label{sec:discussion}

The main result of the present paper is the clear X-ray detection of a
shock front in the intracluster gas of A\,754, 
which coincides with the edge of non-thermal radio diffuse emission at 325 MHz.

Particle acceleration at the shock front cannot be responsible for the whole
Mpc-scale diffuse emission seen in Fig.\ \ref{fig:a754_multifreq}.  Given
the shock parameters derived in \S\ \ref{sec:t_profile}, the downstream velocity 
of the gas is $\simeq 1100$ km~s$^{-1}$, which implies 
that the size of the diffuse emission produced by the shock 
can be only $\sim 100-200$ kpc, once the
radiative lifetime of the emitting electrons ($1-2 \times 10^8$ yr) is taken
into account (see also the discussion in Markevitch et al.\ 2005 and
Brunetti et al.\ 2008). 
However, the spatial coincidence of the ultra-steep
spectrum radio edge with the shock front suggests a physical
connection between these two features, and possibly an indirect
connection between the shock and the diffuse emission on the larger scale. 
If one assumes that relativistic electrons in the relic region 
are accelerated via Fermi mechanism by the shock front with  
$M \sim 1.6$ (as observed in X-rays, see \S\/\ref{sec:t_profile}), 
the expected slope of the radio spectrum from that region, 
accounting for the downstream losses, is $\alpha \sim$ 2.3 
($\alpha = \alpha_{inj} + 1/2$, where $\alpha_{inj} = (p-1)/2$ and 
$p = 2 (M^2 +1)/(M^2 -1)$; e.g., Blandford\& Eichler 1987), 
in rough agreement with the radio observation (see \S\/\ref{sec:radio}).  
However, we can rule out direct shock acceleration on energy grounds. 
If we extrapolate the power law spectrum of electrons responsible for the 
relic emission down to thermal energies 
(as we should if the seed electrons are thermal), the energy in 
relativistic electrons will be of the same order as the thermal energy in the ICM 
requiring an implausible high acceleration efficiency. 
This energy argument implies that  observable radio emission cannot 
be produced at shocks with $M < 2$ (Hoeft \& Brr{\"u}ggen\ 2007). 

Below we discuss possible scenarios for the origin of the cluster-scale
diffuse emission in A754.

\begin{enumerate}  
  \setlength{\itemsep}{1pt}
  \setlength{\parskip}{1pt}
  \setlength{\parsep}{1pt}  
\item  
The edge emission might arise from relativistic plasma that is re-energized
at the shock.  If fossil electrons are present in the ICM, compression
 by the shock can significantly increase their
synchrotron emission at the observing frequency (e.g., Ensslin et al.\ 1998,
Ensslin \& Gopal-Krishna 2001, Markevitch et al 2005).  Adiabatic
compression increases the maximum synchrotron frequency emitted by a
population of fossil electrons by $r^{4/3}$ times (e.g., Markevitch et al. 
2005), where $r$\/ is the shock compression factor (considering a shock
moving perpendicular to the line of sight). In our case, this implies only a
moderate boosting of a factor 2. 
The resulting synchrotron emission may 
light up at our observing frequency.  
The drawback of this scenario is that in this case 
the fossil plasma must have been injected in the ICM
only a few $10^8$ yrs ago, because the fossil electrons must 
exist at energies just below those necessary to emit at the observed frequency, 
and at those energies the lifetime is short. \\
This problem remains even if one assumes that the fossil plasma is confined 
(not mixed) by the ICM (Ensslin \& Gopal-Krishna 2001), 
due to the low Mach number of the shock. 

\item 
Shock reacceleration of fossil relativistic plasma can potentially be an
efficient process even in the case of weak shocks, though this process  
at weak shocks is poorly understood and its efficiency is uncertain. 
Assuming the standard linear shock acceleration theory (e.g., Blandford
\& Eichler 1987), the spectrum of reaccelerated electrons is: 
\begin{equation}
N(\gamma) = (p+2)\gamma^{-p} \int_{\gamma_{min}}^{\gamma}
N_{fss}(x) x^{p-1} dx
\label{reacceleration}
\end{equation}
where $p = 2 (M^2 +1)/(M^2 -1)$ is the slope of the spectrum of accelerated
electrons immediately after the shock, 
$N_{fss}$ is the spectrum of fossil 
electrons in the upstream region, and $\gamma_{min}$ is the minimum 
energy of the fossil electrons at which the reacceleration process works. 
If the fossil relativistic electrons have an exponential cutoff in their power spectrum 
(as they should if they result from radiative cooling, e.g. Sarazin\ 2002),  
shock reacceleration should create a spectrum above that cutoff with a slope similar to 
the one in the classical Fermi reacceleration. 
Assuming $M\sim1.6$, consistent with our findings in \S\ \ref{sec:d_profile}, 
a power-law tail of radio emission with $\alpha \sim 2.3$ 
can be generated at energies $\gamma > 1000$ (Figure\ \ref{fig:reaccel}). 
This scenario avoids the previously mentioned energy argument, because the 
resulting steep spectrum of electrons does not extend down to the thermal energies.

\begin{figure}[htbp]
\epsscale{1.1}
\plotone{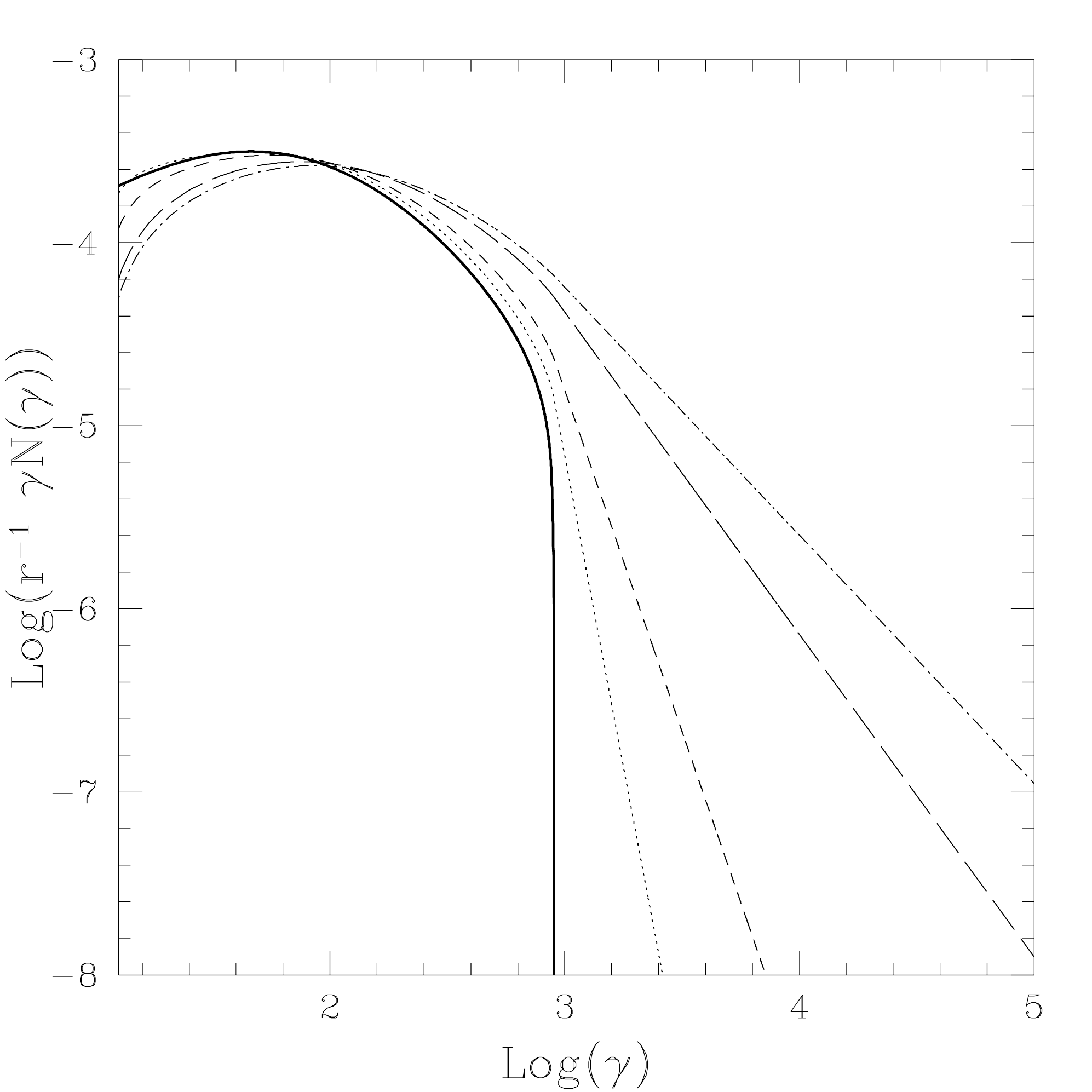}
\caption{
Spectrum of reaccelerated electrons according to Eq.\ \ref{reacceleration} 
We also correct for the shock compression factor in the downstream region, $r$. 
The thick solid line is the initial spectrum of fossil electrons (assuming 
an age of the population of a few Gyrs and typical cluster physical
parameters, e.g. Sarazin\ 2002).
Different models show the spectrum for Mach numbers 1.3 (dotted line),
1.57 (dashed line), 2.5 (long--dashed line), and 3.5 (dot--dashed
line).
A minimum energy $\gamma_{min}=10$ is adopted in the calculations.
}
\label{fig:reaccel}
\end{figure}

\item The connection between the edge and the shock might be indirect.  For
instance, the passage of the shock may have driven small-scale turbulence in
the ICM, which may be long-lived and also produce the radio halo on the larger
scale.  In this case, the radio edge would mark the region where turbulent
acceleration is just beginning to occur. This should happen after $\sim$1
eddy turnover-time of the turbulence, $\approx L/V_l$ (where $L$ and $V_l$ 
are the maximum scale and velocity of the turbulent eddies), since the turbulent
modes at smaller scales are the most important in the particle acceleration
process (e.g., Brunetti \& Lazarian 2007 and references therein). 
This may also explain the
very steep spectrum of the radio emission in the shock region. A possible
indirect connection between shocks and the cluster-wide diffuse radio
emission is suggested by the present radio observation, as well as those of
several other clusters, where bridges of emission connecting radio relics
and halos in several clusters are seen (see, for instance, the discussion in
Brunetti et al.\ 2008). 
A theoretical exploration of this scenario deserves a future effort.
\end{enumerate}


\section{Conclusions}

We have presented a combined X--ray and radio study of the 
nearby merging galaxy cluster Abell\, 754.
\\
The new \chandra\ observation confirms the presence of a 
merger shock front east of the cluster core by providing the 
Mach number from the density jump across the shock  M$=1.57^{+0.16}_{-0.12}$,  
and  a direct measurement of a gas temperature jump of 
the right sign across the arc-like brightness edge, $T_2/T_1 = 2.2^{+1.1}_{-0.7}$.\\
The new \gmrt\, radio image of the cluster at 330 MHz 
reveals that the centrally located diffuse emission 
is very extended and complex, and is mainly elongated in the east-west direction.  
Most interestingly, the eastern edge of this source coincides with the shock.
We studied the spectral properties of the diffuse radio emission,  
using the \gmrt\, data, archival \vla\, data at 1.4 GHz, and previous \vla\, results at 74 MHz. 
We find that the region next to the radio edge (the \textit{relic} region in K01) 
has a very steep integrated spectrum, with $\alpha^{1.4GHz}_{0.3GHz} =  2.02\pm0.04$ and  
$\alpha^{0.3 GHz}_{0.07 GHz} = 1.77\pm0.10$. 
This is steeper than the average spectrum of the whole diffuse emission,   
$\alpha^{1.4 GHz}_{0.3 GHz}$ = 1.57 $\pm$ 0.05, which is in itself steeper than the 
typical spectrum of radio halos. 

The spatial coincidence of the steep spectrum radio edge and the 
merger shock front suggests a physical connection.  
The low Mach number of the shock and the very steep synchrotron spectrum of the edge allow us to 
rule out the scenario of direct shock acceleration of thermal particles for the origin of the edge, 
because it would require an implausible high acceleration efficiency (of order 1). \\
Possible alternative scenarios are reacceleration and/or adiabatic compression 
of fossil relativistic plasma in the ICM. 
Of these, shock reacceleration is more plausible for this weak shock.
It should produce a spectrum of electrons that is in agreement with the 
observed steep radio spectrum of the relic region. \\
Finally, an indirect connection between the shock and the radio edge is also possible. 
The shock passage may drive small-scale turbulence 
that may reaccelerate electrons over the cluster volume.  

Upcoming \gmrt\ deep low-frequency observations will allow us to perform a more 
detailed study of the diffuse emission and help to determine the most plausible scenario 
for its origin. 

\acknowledgements

GM thanks the hospitality of the Harvard-Smithsonian Center for Astrophysics
where most of this work was done.  
We thank N. Kassim and T. Clarke for providing the \vla\, image at 74 MHz  
and for useful comments. 
We thank Alexey Vikhlinin for stimulating discussions. 
We thank the staff of the GMRT for their help during the
observations. GMRT is run by the National Centre for Radio Astrophysics of
the Tata Institute of Fundamental Research. 
This research is partially funded by
INAF under grant PRIN--INAF 2008, and by ASI-INAF under grant E/088/06/0. 
Support was also provided by Chandra 
grant GO8-9128X and NASA contract NAS8-39073.
\\

\end{document}